\documentclass[preprintnumbers,prd,nofootinbib,floatfix]{revtex4}

\pdfoutput=1
\usepackage{hyperref}
\usepackage{graphicx}
\usepackage{dcolumn}
\usepackage{bm}
\usepackage{slashed}

\begin{document}

\title{Dark Matter and Dark Radiation}
\author{Lotty Ackerman, Matthew R. Buckley, Sean M. Carroll, 
and Marc Kamionkowski} 
\affiliation{California Institute of Technology,~Pasadena, CA 91125, USA}
\date{\today}

\begin{abstract}
We explore the feasibility and astrophysical consequences of a new
long-range $U(1)$ gauge field (``dark electromagnetism'') that couples
only to dark matter, not to the Standard Model.  The dark matter consists
of an equal number of positive and negative charges under the new
force, but annihilations are suppressed if the dark matter mass is
sufficiently high and the dark fine-structure constant $\hat\alpha$ is sufficiently
small.  The correct relic abundance can be obtained if the dark matter also
couples to the conventional weak interactions, and we verify that this is
consistent with particle-physics constraints.
The primary limit on $\hat\alpha$ comes from the demand that
the dark matter be effectively collisionless in galactic dynamics, which
implies $\hat\alpha \lesssim 10^{-3}$ for TeV-scale dark matter.  These values 
are easily compatible with constraints from structure formation and primordial
nucleosynthesis.  We raise the prospect of interesting new plasma effects
in dark matter dynamics, which remain to be explored.
 \end{abstract}
\preprint{CALT-68-2704}
\maketitle

\section{Introduction \label{sec:intro}}

A wide variety of cosmological observations seem to point to a
two-component dark sector, in which approximately 73\% of the
energy density of the universe is in dark energy and 23\% is in 
non-baryonic dark matter (DM). Ordinary matter constitutes the remaining 4\%
\cite{Amsler:2008zz}.  The physics of the dark matter sector is plausibly quite minimal:
an excellent fit to the data is obtained by assuming that dark matter
is a cold, collisionless relic, with only the relic abundance as a free parameter.  The well-known
``WIMP miracle'' \cite{Jungman:1995df,Bergstrom:2000pn,Bertone:2004pz} is the fact that a stable, neutral particle with weak-scale 
mass and coupling naturally provides a reasonable energy density in DM.
Particles of this type arise in models of low-scale supersymmetry \cite{Jungman:1995df} or
large extra dimensions \cite{Hooper:2007qk}, and provide compelling DM candidates.
In the contemporary universe, they would be collisionless as far as
any conceivable dynamical effects are concerned.

Nevertheless, it is also possible to imagine a rich phenomenology
within the dark sector.  The dark matter could be
coupled to a relatively strong short-range force that could have interesting
consequences for structure on small scales \cite{Spergel:1999mh,Wandelt:2000ad}.
Alternatively, DM could also be weakly coupled to long-range forces, which might
be related to dark energy \cite{Gradwohl:1992ue}.  
One difficulty with the latter is that such forces are
typically mediated by scalar fields, and it is very hard to construct natural
models in which the scalar field remains massless (to provide a long-range
force) while interacting with the DM at an interesting strength.

In this paper, we explore the possibility of a long-range \emph{gauge} 
force coupled to DM, in the form of a new unbroken abelian field,
dubbed the $U(1)_D$ ``dark photon.''  We imagine that this new gauge boson $\hat\gamma$
couples to a DM fermion $\chi$, but not directly to any Standard Model (SM) 
fields.  Our model is effectively parameterized by only two numbers:  $m_\chi$, the
mass of the DM, and $\hat\alpha$, the dark fine-structure constant.  If
$m_\chi$ is sufficiently large and $\hat\alpha$ is sufficiently small, annihilations
of DM particles through the new force
freeze out in the early universe and are negligible today, despite there being
equal numbers of positively- and negatively-charged particles.
The dark matter in our model is therefore a plasma, which could conceivably
lead to interesting collective effects in the DM dynamics.

Remarkably, the allowed values of $m_\chi$ and $\hat\alpha$ seem quite
reasonable.  We find that the most relevant constraint comes from demanding that
accumulated soft scatterings do not appreciably perturb the motion of DM particles
in a galaxy over the lifetime of the universe, which can be satisfied by 
$\hat \alpha \sim 10^{-3}$ and $m_\chi \sim$~TeV. For values near these bounds, the alterations in DM halo shapes may in fact lead to closer agreement with observation \cite{Spergel:1999mh}.
However, for such regions of parameter space, if $U(1)_D$  were the only interaction felt
by the $\chi$ particles, the resulting relic abundances would be slightly too large, so we
need to invoke an additional annihilation channel.  We show that $\chi$ can 
in fact be a WIMP, possessing $SU(2)_L$ quantum numbers in addition to
$U(1)_D$ charge.  Such a model provides the correct relic abundance, and 
is consistent with particle-physics constraints so long as the mixing between ordinary
photons and dark photons is sufficiently small.

We consider a number of other possible observational limits on dark electromagnetism, and
show that they do not appreciably constrain the parameter space.  Since the DM halo is
overall neutral under $U(1)_D$, there is no net long-range force that violates the equivalence principle.
Although there are new light degrees of freedom, their temperature is naturally lower
than that of the SM plasma, thereby avoiding constraints from Big-Bang Nucleosynthesis (BBN).
Energy loss through dark bremsstrahlung radiation is less important than the 
soft-scattering effects already mentioned.
The coupling of DM to the dark radiation background can in principle suppress the
growth of structure on small scales, but we show that the DM decouples from the 
dark radiation at an extremely high redshift.  On the other hand, we find that there are
plasma instabilities ({\it e.g.}~the Weibel instability) that can potentially play an
important role in the assembly of galactic halos; however, a detailed analysis of these
effects is beyond the scope of this work.

The idea of an unbroken $U(1)$ coupled to dark matter is not new.\footnote{Broken $U(1)$ forces have, of course, also been considered, see {\it e.g.}~Ref.~\cite{Hooper:2008im}}
De~Rujula et al. \cite{DeRujula:1989fe} explored the possibility that dark matter was
charged under conventional electromagnetism (see also \cite{Dimopoulos:1989hk,
Holdom:1985ag,Davidson:2000hf,Chuzhoy:2008zy}).  Gubser and Peebles
\cite{Gubser:2004uh} considered structure formation in the presence of both
scalar and gauge long-range forces, but concentrated on a region of parameter space
in which the gauge fields were subdominant.  Refs. \cite{Feng:2008ya,Feng:2008mu} considered several models for a hidden dark sector, including one manifestation in which the dark matter consists of heavy hidden-sector staus interacting via a copy of electromagnetism. The effect of dimension-6 operators containing a new $U(1)$ gauge boson and SM fields was considered in Ref.~\cite{Dobrescu:2004wz}, for models where the only fields in a hidden sector are charged under the new force. Additional models which contain unbroken abelian gauge groups may be found, for example in Refs.~\cite{Pospelov:2007mp,Ahluwalia:2007cq}.
In this paper, we construct minimal models of
dark matter coupled to a new unbroken $U(1)_D$, leaving the dark fine-structure
constant and dark-matter mass as free parameters, and explore what regions of 
parameter space are consistent with astrophysical observations and what new
phenomena might arise via the long-range gauge interaction.

In Section~\ref{sec:earlyuniverse}, we introduce our notation for a minimal dark-matter sector including a new abelian symmetry $U(1)_D$. We then consider the bounds on the new dark parameters from successful thermal production of sufficient quantities of dark matter as well as requiring that BBN and cosmic microwave background (CMB) predictions remain unchanged. The restrictions of parameter space are closely related to those resulting from standard short-range WIMP dark matter. In Section~\ref{sec:galacticdynamics}, we consider the effect of long range interactions on DM particle interactions in the halos of galaxies. By requiring that our model not deviate too greatly from the predictions of collisionless DM, we find that the allowed regions of $\hat{\alpha}/m_\chi$ parameter space from Section~\ref{sec:earlyuniverse} are essentially excluded. In order to evade these constraints, Section~\ref{sec:SU2} describes an extended model, where the dark-matter candidate is charged under both $SU(2)_L$ and the new $U(1)_D$. Additional effects of dark radiation are presented in Section~\ref{sec:newlimits}, and we conclude in Section~\ref{sec:conclusion}.

We note that our model does not address the hierarchy problem, nor provide a high-energy 
completion to the SM. However, new gauge groups and hidden sectors may be generic results of many such high-energy theories ({\it e.g.}~string and grand unified theories),
and a WIMP coupled to an unbroken $U(1)$ is certainly a plausible low-energy
manifestation of such theories. 
The most important lesson of our model is that interesting physics might be lurking
in the dark sector, and it is worthwhile to consider a variety of possible models and
explore their consequences for astrophysics and particle physics.

\section{Dark Radiation and the Early Universe \label{sec:earlyuniverse}}

We postulate a new ``dark'' abelian gauge group $U(1)_D$ with gauge coupling constant $\hat{g}$ 
and dark fine-structure constant $\hat{\alpha} \equiv \hat{g}^2/4\pi$. In the simplest case, the dark matter sector consists of a single particle $\chi$ 
with $U(1)_D$ charge of $+1$ and mass $m_\chi$ along with its antiparticle $\bar{\chi}$. For definiteness, we take $\chi$ to be a fermion, though 
 our results are applicable to scalars as well. As the limits on new long range forces on SM fields are very stringent, we assume that all the SM fields are neutral under $U(1)_D$. 
For the moment we take the $\chi$ field to be a singlet under $SU(3)_C\times SU(2)_L\times U(1)_Y$, a restriction that will be relaxed in Section~\ref{sec:SU2}. As a result, this extension of the SM is anomaly free. In this Section, we will derive constraints on the mass $m_\chi$ and coupling $\hat{\alpha}$ from the evolution of dark matter in the early universe. Two considerations drive these constraints: the dark matter must provide the right relic abundance at thermal freeze-out, and the dark radiation from the $U(1)_D$ cannot contribute too greatly to relativistic degrees of freedom at BBN (a similar bound coming from the CMB also applies but is weaker).

The degrees of freedom in the dark sector are thus the heavy DM fermions $\chi$ and massless dark photons $\hat{\gamma}$. The Lagrangian for the dark sector is
\begin{equation}
{\cal L} =  \bar{\chi}(i \slashed{D}+m_\chi) \chi - \frac{1}{4} \hat{F}_{\mu\nu}\hat{F}^{\mu\nu}. \label{eq:lag}
\end{equation}
Here $D_\mu = \partial_\mu - i\hat{g} \hat{A}_\mu$ and $\hat{F}_{\mu\nu}$ is the field-strength tensor for the dark photons. We assume that the mixing term $c\hat{F}_{\mu\nu}F^{\mu\nu}$ is set to zero at some high scale (say the GUT scale).
This is a self-consistent choice, since if there is no mixing between the dark and visible sectors, $c=0$ is preserved by the renormalization group evolution.
(In Section~\ref{sec:SU2} we argue that mixing is not generated by radiative corrections even when $\chi$ carries $SU(2)_L$ quantum numbers.) This choice allows us to bypass constraints on a new $U(1)$ coming from mixing between the photon and dark photon, that is, `paraphotons' \cite{Okun:1982xi,Holdom:1985ag}.  We have no {\it a priori} assumptions on the parameters $m_\chi$ and $\hat{\alpha}$, though as we shall see, it suffices to think of the former as ${\cal O}(100-1000~\mbox{GeV})$ and the latter $\lesssim {\cal O}(10^{-2})$. 

We now follow the thermal history of the dark sector. Our analysis follows that of Ref.~\cite{Feng:2008mu}; we rehearse it in a slightly simpler context here to illustrate how the results depend on our various assumptions. If the visible sector and the dark sector are decoupled from
each other, they may have different temperatures, $T$ and $\hat T$, respectively; rapid
interactions between them would equilibrate these two values.  After inflation,
the two sectors could conceivably reheat to different temperatures, depending on the coupling of
the inflaton to the various fields.  Even if the temperatures are initially equal, once the two sectors decouple as the universe expands and cools, entropy deposited from frozen-out degrees of freedom in one sector will generally prevent the dark temperature $\hat{T}$ from tracking the visible sector temperature $T$. The ratio 
\begin{equation}
  \xi = \hat{T}/T
\end{equation}
will depend on the spectrum of both sectors, and is itself a function of $T$. 

As the temperature drops below a particle's mass, the associated degrees of freedom freeze out and dump entropy into their respective sectors (dark or visible). This causes the temperature of that sector to decline more slowly than $1/a$, where $a$ is the scale factor of the universe. As the entropy density $s$ of the visible sector and $\hat{s}$ of the dark sector are individually conserved after decoupling, we must separately count the degrees of freedom in these two sectors. There are two definitions of degrees of freedom of interest to us: $g_*$ and $g_{*\rm S}$. The former is defined as
\begin{equation}
g_* = \sum_{i= {\rm bosons}} g_i \left(\frac{T_i}{T}\right)^4 +  \frac{7}{8} \sum_{i= {\rm fermions}} g_i \left(\frac{T_i}{T}\right)^4, \label{eq:gstar}
\end{equation}
and is used in calculation of the total relativistic energy density, $\rho_R \propto g_* T^4$. Here $g_i$ is the number of degree of freedom for particle species $i$, $T_i$ is the temperature of the thermal bath of species $i$, and $T$ is the temperature of the photon bath. The sums run over all active degrees of freedom at temperature $T$. Separating out the visible fields, $g_*$ can be written as
\begin{equation}
g_* = g_{*{\rm vis}}+ \sum_{i= {\rm bosons}} g_i \xi(T)^4 +  \frac{7}{8} \sum_{i= {\rm fermions}} g_i \xi(T)^4 \label{eq:gstarxi}
\end{equation}
where the sums now run over the dark particles. If we restrict the visible sector to the SM, then the term $g_{*{\rm vis}}$ is $106.75$ above the top mass, dropping gradually to $\sim 60$ at $T=\Lambda_{\rm QCD}$. Between $100~\mbox{MeV} \gtrsim T \gtrsim 1~\mbox{MeV}$, $g_{*{\rm vis}}=10.75$, and drops again to $3.36$ in the present day. (See {\it e.g.}~Ref.~\cite{KolbTurner} for more detail.)

Similarly, the total entropy density $s_{\rm tot}$ (a conserved quantity) at a photon temperature $T$ is proportional to $g_{*\rm{S}} T^3$, where
\begin{eqnarray}
g_{*{\rm S}} & = & \sum_{i= {\rm bosons}} g_i \left(\frac{T_i}{T}\right)^3 +  \frac{7}{8} \sum_{i= {\rm fermions}} g_i \left(\frac{T_i}{T}\right)^3 \label{eq:gstarS} \\
 & = & g_{*{\rm S,vis}}+ \sum_{i= {\rm bosons}} g_i \xi(T)^3 +  \frac{7}{8} \sum_{i= {\rm fermions}} g_i \xi(T)^3\,. 
 \label{eq:gstarSxi}
\end{eqnarray}
Prior to neutrino decoupling, all the relativistic standard model degrees of freedom are in thermal equilibrium at a common temperature. Thus, before $T\sim 1$~MeV when neutrinos decouple, we have $g_{*{\rm vis}}=g_{*{\rm S,vis}}$. Furthermore, we may split the dark $g_{*\rm S}$ into heavy and light degrees of freedom: $g_{\rm heavy}$ and $g_{\rm light}$, where the heavy degrees of freedom are non-relativistic at BBN. We are interested in the number of degrees of freedom at BBN ($T\sim 1$~MeV) because formation of the experimentally observed ratios of nuclei are very sensitive to the expansion of the universe at that time, which is related to the energy density of radiation through the Friedmann equation. From this, a bound on the number of relativistic degrees of freedom can be derived \cite{Feng:2008mu}.

\begin{figure}[ht]
\centering

\includegraphics[width=0.5\columnwidth]{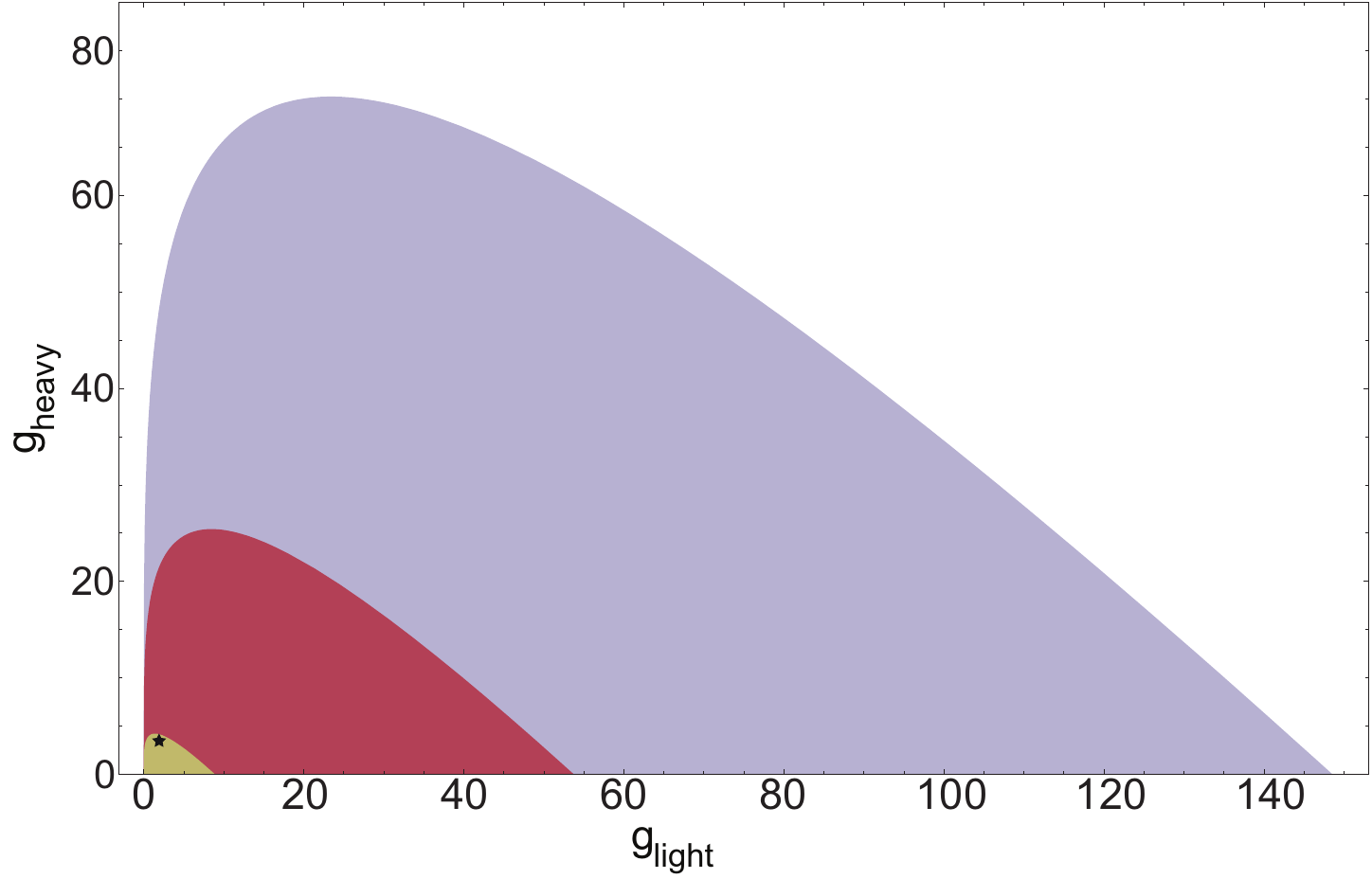}

\caption{The allowed values of dark $g_{\rm light}$ (those degrees of freedom relativistic at $T_{\rm BBN}$ ) and $g_{\rm heavy}$ (the remaining dark degrees of freedom) arising from BBN constraints Eqs.~(\ref{eq:BBNbound}) and (\ref{eq:BBNbound2}). The allowed regions correspond to 95\% confidence levels for $\xi(T_{\rm RH})=1$ and a visible sector $g_{*\rm vis} = 106.75$ (red),  $\xi(T_{\rm RH})=1$ and $g_{*\rm vis} =228.75$ (corresponding to MSSM particle content, in blue), and $\xi(T_{\rm RH})=1.4(1.7)$ and $g_{*\rm vis} = 106.75(228.75)$ (in yellow). The minimal dark sector model of this paper is noted by a black star at $g_{\rm light} = 2$ and $g_{\rm heavy} = 3.5$. \label{fig:gstar}}
\end{figure}

Using the separate conservation of the visible and dark sector entropy and the previous definitions, we see that, at BBN
\begin{equation}
\frac{g_{\rm light} \xi(T_{\rm BBN})^3}{(g_{\rm heavy}+g_{\rm light})\xi(T_{\rm RH})^3} = \frac{g_{*\rm{vis}}(T_{\rm BBN})}{g_{*{\rm vis}}(T_{\rm RH})} \label{eq:gstarBBN}
\end{equation}
here we have set $g_{*{\rm S,vis}} = g_{*\rm vis}$ (recall that $g_{*\rm{vis}}(T_{\rm BBN}) = 10.75$). The BBN bound on relativistic degrees of freedom is usually stated in terms of number of light neutrino species in thermal equilibrium at the time: $N_\nu = 3.24 \pm 1.2$ \cite{Cyburt:2004yc}. Here the error bars correspond to $2\sigma$ (95\% confidence). Therefore, assuming three light neutrino species in the visible sector, if the dark sector is not to violate this bound, we must require
\begin{equation}
g_{\rm light} \xi(T_{\rm BBN})^4 = \frac{7}{8}\times 2 \times (N_\nu-3) \leq 2.52 ~~(95\%~\mbox{confidence}). \label{eq:BBNbound}
\end{equation}
Combining Eqs.~(\ref{eq:gstarBBN}) and (\ref{eq:BBNbound}), we find that
\begin{equation}
g_{\rm light}\left[\frac{g_{\rm heavy}+g_{\rm light}}{g_{\rm light}}\frac{10.75}{g_{*{\rm vis}}(T_{\rm RH})} \right]^{4/3} \xi(T_{\rm RH})^4 \leq 2.52 ~~(95\%~\mbox{confidence}). \label{eq:BBNbound2}
\end{equation}
Since the high energy completion of the visible sector must at minimum include the SM fields, $g_{*\rm vis}(T_{\rm RH}) \geq 106.75$; a bound on the dark sector $g_{\rm light}$ and $g_{\rm heavy}$ can be derived for a fixed value of $\xi(T_{\rm RH})$ (see Fig.~\ref{fig:gstar}). Increasing the number of visible sector degrees of freedom at high temperatures (for example to that of the MSSM) relaxes this bound. 

In the case of $\xi(T_{\rm RH}) =1 $, we see that the minimal model of the dark sector (only heavy $\chi/\bar{\chi}$ and light $\hat{\gamma}$) is safely included. Due to the fourth power of $\xi$ entering into Eq.~(\ref{eq:BBNbound2}), if the minimal dark sector is not to be ruled out, we find $\xi(T_{\rm RH}) \leq 1.4(1.7)$ for the SM(MSSM) particle content. A similar bound on relativistic degrees of freedom can be derived from the cosmic microwave background, but provides a weaker $2\sigma$ exclusion limit \cite{Smith:2006nka,Feng:2008mu}.

\begin{figure}[t]
\centering

\includegraphics[width=0.25 \columnwidth]{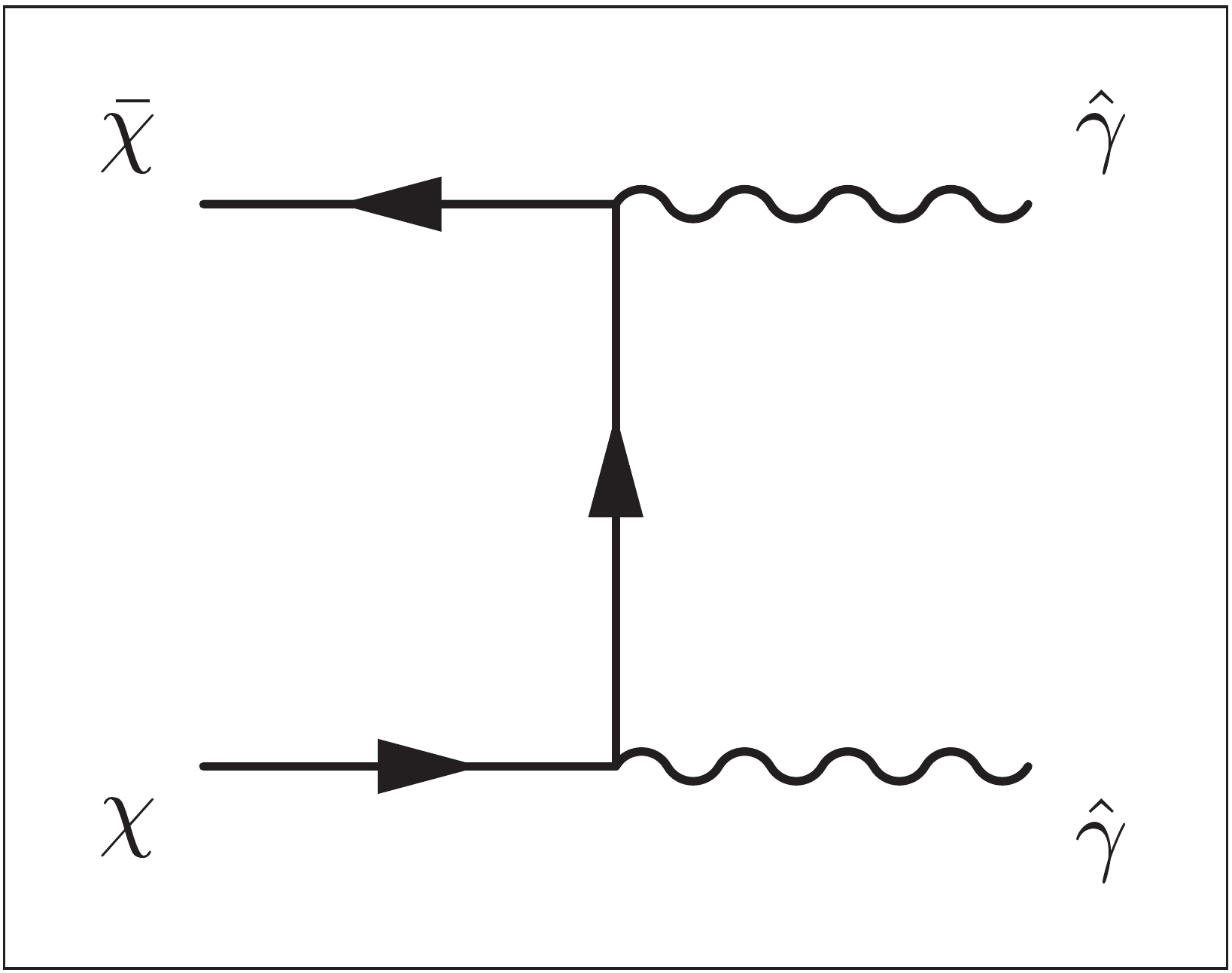} \includegraphics[width=0.25 \columnwidth]{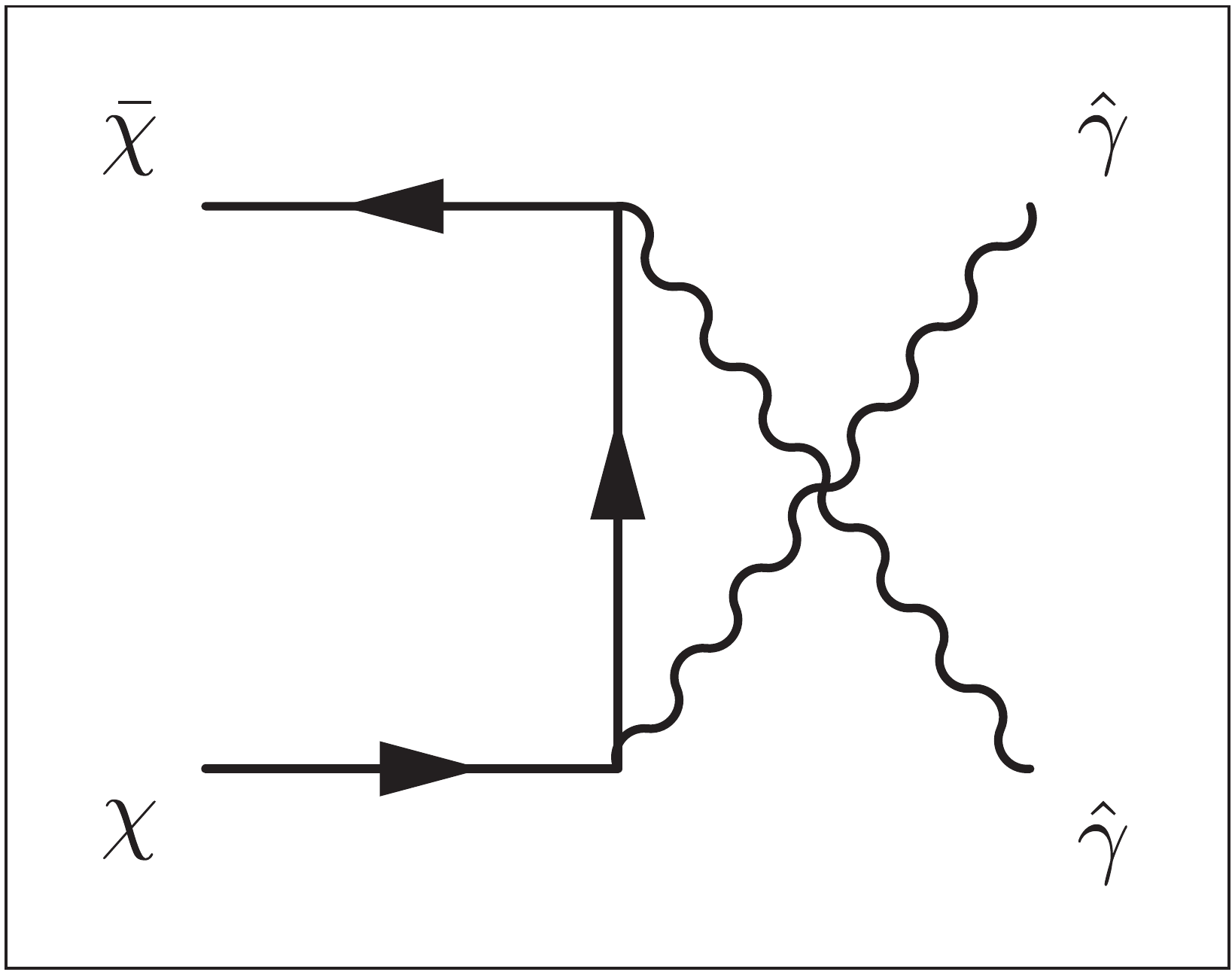}
\caption{Pair annihilation/creation of dark matter $\chi$ into dark photons $\hat{\gamma}$ via $t$ and $u$-channel exchange diagrams. These processes keep the dark sector in thermal equilibrium until the $\chi$ particles become non-relativistic. \label{fig:feynman}}
\end{figure}

We now turn to bounds on the coupling $\hat{\alpha}$ and dark matter mass $m_\chi$ coming from the dark matter abundance. At temperatures $\hat{T}$ much above $m_\chi$, the $\chi$ particles are kept in thermal equilibrium with the dark photons $\hat{\gamma}$ (and possibly other particles in the dark sector) via pair annihilation/creation as in the Feynman diagrams of Fig.~\ref{fig:feynman}. Since the annihilation can proceed via $s$-wave processes, the thermally averaged cross section $\langle \sigma v\rangle$ is, to leading order, independent of $v$:\footnote{Strictly speaking, there will be a Sommerfeld enhancement in this cross section in the limit $v\rightarrow0$ \cite{Hisano:2004ds}.  This will slightly change the relic abundance \cite{Kamionkowski:2008gj}, but we leave the detailed analysis for future work.}
\begin{equation}
\langle \sigma v\rangle \approx \sigma_0 = \frac{\pi \hat{\alpha}^2}{2m_\chi^2} + {\cal O}(v^2).\label{eq:sigma}
\end{equation}
Using this, the relic density of the $\chi$ particles may be easily calculated (see, for example Ref.~\cite{KolbTurner}).

As a rule of thumb, the dark matter drops out of thermal equilibrium when the rate $\Gamma$ of annihilation $\chi\bar{\chi} \to \hat{\gamma}\hat{\gamma}$ (and the reverse process) is outpaced by the expansion of the universe $H$. Using the Boltzmann equation, the contribution of $\chi$ to the energy density of the universe can be more precisely calculated as
\begin{equation}
\Omega_{\rm DM} h^2 = 1.07 \times 10^9 \frac{(n+1)x_f^{n+1}~\mbox{GeV}^{-1}}{ (g_{*\rm{S}}/\sqrt{g_*}) m_{\rm Pl}\sigma_0}. \label{eq:OmegaDM}
\end{equation}
Here $x_f$ is the ratio $m_{\chi}/\hat{T}_f$ where $\hat{T}_f$ is the dark temperature at time of freeze-out and $n=0$ for $s$-wave processes. The quantity $x_f$ is given by
\begin{equation}
x_f = \ln\left[ 0.038(n+1)\left(\frac{g}{\sqrt{g_*}} \right) m_{\rm Pl} m_\chi \sigma_0 \right]-\left(n+\frac{1}{2}\right)\ln\ln\left[ 0.038(n+1)\left(\frac{g}{\sqrt{g_*}} \right) m_{\rm Pl} m_\chi \sigma_0 \right],\label{eq:xf}
\end{equation}
where $g$ is the number of degrees of freedom in the $\chi$ system (namely 4).

As $g_*$ enters into the formula for $x_f$ only logarithmically, we may make the approximation that $g_{*\rm S} \approx 100$ if $\chi$ freezes out while $T$ is above $\Lambda_{\rm QCD}$. We make the additional assumptions that the only degrees of freedom in addition to the SM are the $\hat{\gamma}$ and $\chi$ in the dark sector and that $\xi(T_{\rm RH})=1$. We shall consider how these assumptions may be relaxed later. 

\begin{figure}[htbp]
\centering

\includegraphics[width=0.8\columnwidth]{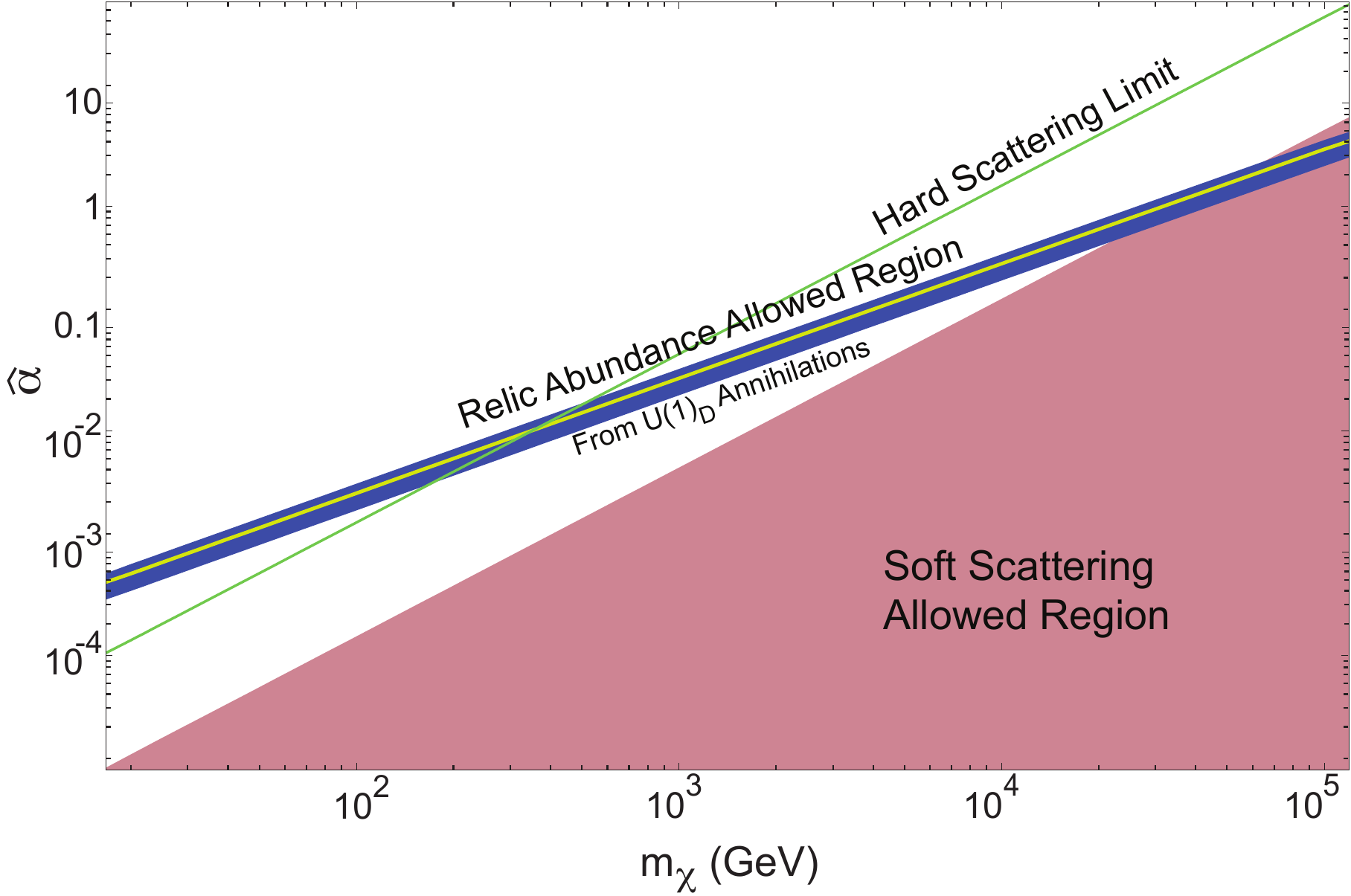}

\caption{The allowed regions of $\hat{\alpha}$ vs. $m_\chi$ parameter space.  The relic 
abundance allowed region applies to models in which $U(1)_D$ is the only force
coupled to the dark matter; in models where the DM is also weakly interacting, this
provides only an upper limit on $\hat\alpha$.  The thin yellow line is the allowed region from 
correct relic abundance assuming $\Omega_{\rm DM}h^2 = 0.106\pm 0.08$, 
$\xi(T_{\rm RH}) = 1$, $g_{*\rm vis} \approx 100$, and $g_{\rm heavy}+g_{\rm light}= 5.5$
while the surrounding blue region is $g_{*\rm vis} = 228.75(60)$, $\xi(T_{\rm RH}) = 1(0.1)$, 
and $g_{\rm heavy}+g_{\rm light}= 100(5.5)$ at the lower(upper) edge. 
The diagonal green line is the upper limit on $\hat\alpha$ from effects of hard scattering on
galactic dynamics; in the red region, even soft scatterings do not appreciably affect the
DM dynamics.  We consider this to be the allowed region of parameter space.
\label{fig:alphamass}}
\end{figure}

Under these assumptions, the contribution of the dark sector to $g_{*}$ and $g_{*\rm S}$ is $2+(7/8)\times 4= 11/2$. As no dark degrees of freedom have frozen out yet, $\xi(T_f) = \left( \frac{g_{*\rm vis}(T_f)}{g_{*\rm vis}(T_{\rm RH})}\right)^{1/3} \xi(T_{\rm RH}) \approx 1$. With the measured value $\Omega_{DM}h^2 = 0.106\pm 0.08$ \cite{Amsler:2008zz},
we may solve for the allowed values of $\hat{\alpha}$ as a function of $m_\chi$ in Eq.~(\ref{eq:OmegaDM}). The resulting band is shown in Fig.~\ref{fig:alphamass}.

In this discussion we have assumed that the process which sets the relic
abundance of $\chi$ is annihilation into $\hat\gamma$s, as shown in Figure~{\ref{fig:feynman}}.
As we will argue in the next section (and as is already shown in Figure~\ref{fig:alphamass}), 
the values we obtain for $\hat\alpha$ from this
calculation are incompatible with bounds from galactic dynamics unless $m_\chi > 10^5$~GeV (at which point $\hat{\alpha}$ is non-perturbative). 
However, we can get the correct relic abundance even with much lower values of $\hat\alpha$
by adding other annihilation channels, such as the weak interactions, as explored in
Section~\ref{sec:SU2}.  In that case, the ``relic abundance allowed region'' discussed here
really becomes an upper limit; if the dark fine-structure constant is larger than that value,
annihilations are too efficient, and the correct abundance cannot be obtained.

We now consider how changing our assumptions on $g_{*}$ and $\xi$ can change our conclusions on the allowed parameter space. The parameter $\xi(T_f)$ does not enter explicitly into the calculation for $\Omega_{\rm DM}h^2$, however it does affect the number of active degrees of freedom at freeze-out directly, through Eqs.~(\ref{eq:gstarxi}) and (\ref{eq:gstarSxi}), and indirectly by allowing the temperature $T$ to differ from $\hat{T}$.
If $\xi < 1$, $\hat{T} < T$ and there could be many more heavy visible degrees of freedom still active when $\chi$ freezes out. $\xi > 1$ would reduce the visible degrees of freedom. However, as we have seen in Eq.~(\ref{eq:BBNbound2}), it is difficult to construct a scenario with large $\xi$, short of a massive increase in $g_{*\rm vis}$ and small values of $g_{*\rm heavy}+g_{*\rm light}$. We include in Fig.~\ref{fig:alphamass} the bounds from both a large and small value of $g_{*}$. The large limit is $g_{*\rm vis}(T_f) = 228.75$, ({\it i.e.}~equivalent to the MSSM degrees of freedom), $\xi(T_{\rm RH}) = 1$, and $g_{\rm heavy}+g_{\rm light} = 100$, while the small value is given by $g_{*\rm vis}(T_f) = 60$, ({\it i.e.}~equivalent to the SM degrees of freedom at $\Lambda_{\rm QCD}$), $\xi(T_{\rm RH}) = 0.1$, and $g_{\rm heavy}+g_{\rm light} = 5.5$.

\section{Galactic Dynamics \label{sec:galacticdynamics}}

Although freezeout in our scenario is similar to that in the
standard WIMP scenario, the long-range DM-DM interactions implied by
the unbroken $U(1)_D$ may lead to considerably different DM
phenomenology in the current Universe, and in particular in
galactic halos.  In this scenario, dark-matter halos are
composed of an equal mixture of $\chi$ and $\bar{\chi}$.  The
overall halo will be $U(1)_D$ neutral, eliminating long-range
forces that are incompatible with experiment.

However, nearest-neighbor interactions between $\chi$
particles remain, and these interactions can be constrained by
observations that suggest that dark matter is effectively
collisionless.  Constraints to dark-matter self-interactions
arise from evidence for nonspherical cores for some dark-matter
halos (collisions tend to make the cores of halos round) \cite{MiraldaEscude:2000qt} and
from evidence for dark-matter halos with large phase-space
densities (collisions would reduce phase-space densities) \cite{Wandelt:2000ad,Dave:2000ar,Yoshida:2000uw}.
Roughly speaking, a bound to DM-DM interactions can be derived
by demanding that scattering induces no more than a small
fractional change in the energy of a typical DM particle in a
galactic halo during the history of the Universe \cite{Spergel:1999mh}. This translates to an upper bound of $\sim 0.1 \mbox{ cm$^2$/g}$ on the more familiar quantity $\sigma/m_\chi$.\footnote{This can be seen from Eq.~(\ref{eq:tau}), using the age of the universe for $\tau$, and Galactic parameters $\rho = n m_\chi = 0.3~\mbox{GeV/cm$^3$}$, $v/c = 10^{-3}$.} A separate bound of $\sigma/m_\chi < 1.25$ can be derived from the Bullet Cluster \cite{Clowe:2006eq,Randall:2007ph}, but as this is less restrictive we ignore it here.

To illustrate, we first consider hard scattering of
a $\chi$ off another $\chi$ or $\bar{\chi}$, where energy on the
order of  $m_\chi v^2/2$ is exchanged. 
The mean free time $\tau$ for a $\chi$ to undergo a hard
scattering with another $\chi(\bar{\chi})$ is given by
\begin{equation}
     \tau = \frac{1}{\langle n \sigma v \rangle}, \label{eq:tau}
\end{equation}
where $n$ is the number density of dark matter, $\sigma$ is the
hard-scattering cross section, and $v$ is the velocity of the
dark-matter particles. The number $N$ of dark-matter particles
in the Galaxy is
\begin{equation}
     N = \frac{M_{\rm Gal}}{m_\chi} \approx 10^{64} \left(
     \frac{m_\chi}{\mbox{TeV}} \right)^{-1}, \label{eq:N}
\end{equation}
and $n \approx 3N/4 \pi R^3$, where $R$ is the radius of the
Galaxy. The velocity $v$ is
\begin{equation}
     v \simeq \sqrt{\frac{G M_{\rm Gal}}{R}} \simeq
     \sqrt{\frac{GNm_\chi}{R}}. \label{eq:v}
\end{equation}
The dynamical time $\tau_{\rm dyn}$ in the Galaxy is
\begin{equation}
\tau_{\rm dyn} = 2\pi R/ v. \label{eq:taudyn}
\end{equation}
Taking $\tau_{\rm dyn} \approx 2 \times 10^{8}$~years for the
Milky Way, the average time for a hard scatter for a dark-matter
particle is greater than the age of the universe if
\begin{equation}
     \frac{\tau}{\tau_{\rm dyn}} = \frac{2R^2}{3N\sigma} \gtrsim
     50. \label{eq:tauconstraint}
\end{equation}

A hard scatter occurs when two particles pass close
enough so that their kinetic energy is comparable to their
potential energy.  The impact parameter
that defines a hard scatter is thus
\begin{equation}
     b_{\rm hard} = \frac{2 \hat{\alpha}}{v^2
     m_\chi}. \label{eq:bhard}
\end{equation}
Taking the cross section for hard scatters to be $\sigma_{\rm
hard} \approx b^2_{\rm hard}$, and using Eq.~(\ref{eq:v}) for
$v$, we find
\begin{equation}
     \frac{\tau_{\rm hard}}{\tau_{\rm dyn}} = \frac{G^2 m_\chi^4 N}{6
     \hat{\alpha}^2} \gtrsim 50. \label{eq:tauhard} 
\end{equation}
Using $G = m_{\rm Pl}^{-2} \approx 10^{-32}~\mbox{TeV}^{-2}$ we
find the hard scattering limit on the $U(1)_D$ coupling constant
to be
\begin{equation}
     \hat{\alpha} \lesssim \sqrt{\frac{1}{300}}
     \left(\frac{m_\chi}{\mbox{TeV}}\right)^{3/2}= 0.06
     \left(\frac{m_\chi}{\mbox{TeV}}\right)^{3/2}. \label{eq:alphahard}
\end{equation}
The allowed region arising from this bound is shown in
Fig.~\ref{fig:alphamass}.

We now turn to the effect of soft scattering on the allowed
values of $\hat{\alpha}$ and $m_\chi$. Here we consider the approach of one $\chi$
particle towards another $\chi(\bar{\chi})$ at impact parameter
$b$. By definition, for soft scattering $b>b_{\rm hard}$. The
velocity change induced by the encounter is
\begin{equation}
     \delta v = \pm \frac{2 \hat{\alpha}}{m_\chi b
     v}. \label{eq:deltav} 
\end{equation}
As one dark-matter particle orbits the Galaxy, it sees a surface
density $N/\pi R^2$ of dark matter. The number of interactions
that occur between an impact parameter $b$ and $db$ is $\delta n
= (N/\pi R^2) 2\pi b db$. While the change in $\delta v$ over
these interactions should average to zero, this is not true for
$\delta v^2$:
\begin{equation}
     \delta v^2 = (\delta v)^2 \delta n =
     \frac{8\hat{\alpha}^2 N}{m_\chi^2 v^2R^2} b^{-1}
     db. \label{eq:deltav2}
\end{equation}
Integrating $\delta v^2$ from $b_{\rm hard}$ to the maximum
possible impact parameter in the Galaxy, $R$, gives the total
change in $v^2$ as the particle orbits once through the halo:
\begin{equation}
     \Delta v^2 = \frac{8\hat{\alpha}^2 N}{m_\chi^2
     v^2R^2}\ln(R/b_{\rm hard}) = \frac{8\hat\alpha^2
     N}{m_\chi^2 v^2R^2}\ln\left( \frac{GNm_\chi^2}{2
     \hat{\alpha}}\right).\label{eq:totaldeltav2}
\end{equation}
The number $\tau/\tau_{\rm dyn}$ of orbits it will take for the
dark-matter particle to have $\Delta v^2/v^2 \sim {\cal O}(1)$ is
\begin{equation}
     \frac{\tau_{\rm soft}}{\tau_{\rm dyn}} = \frac{G^2 m_\chi^4 N}{8
     \hat{\alpha}^2} \ln^{-1}\left(  \frac{GNm_\chi^2}{2
     \hat{\alpha}}\right) \gtrsim 50. \label{eq:tausoft}
\end{equation}
The logarithmic suppression in Eq.~(\ref{eq:tausoft}) relative
to Eq.~(\ref{eq:tauhard}) is due to the long-range Coulomb force
generated by the $U(1)_D$. As can be seen in
Fig.~\ref{fig:alphamass}, the allowed region from these
considerations of Galactic dynamics completely exclude the
$\hat{\alpha}/m_\chi$ band that gives the correct relic
abundance up to $m_\chi \sim 30$~TeV. For $m_\chi\sim 1$~TeV a dark matter candidate
which freezes out due to $U(1)_D$ interactions is ruled out from
such considerations. In particular, models such as that in Ref.~\cite{Feng:2008mu} with $m_\chi \sim m_W$ and a hidden copy of electromagnetism ({\it i.e.}~$\hat{\alpha}=1/137$) are ruled out, even though the freeze-out proceeds through hidden-sector weak interactions rather than a $U(1)_D$. Interestingly, $\hat\alpha = \alpha$ is allowed for $m_\chi \gtrsim 2$~TeV.

Before considering whether such a model may be valid if our assumptions are loosened, we should ask why Galactic dynamics do not similarly exclude WIMP dark matter. After all, both models have similar cross sections for annihilations in the early universe (Eq.~(\ref{eq:sigma})) as is required for the correct relic density. Though the soft scattering limit clearly will not apply due to the short range nature of the broken $SU(2)_L$, naively it would seem that the hard scattering limit Eq.~(\ref{eq:tauhard}) should apply to WIMPs equally well. However, notice that the threshold for hard scattering with a $U(1)_D$ is dependent on energy. As the temperature drops, the cross section rises, as the $\chi$ particles no longer have to approach as close in order for to potential energy $V(r)$ to be of the order of the kinetic energy. Contrast this to hard scattering from WIMPs, where the cross section is always proportional to $\alpha^2/m_{\rm DM}^2$, regardless of the velocity. Entering this cross section into Eq.~(\ref{eq:tauconstraint}), results in the uninteresting bound that $m_{\rm DM} \lesssim 10^{13}$~TeV for WIMP dark matter from Galactic dynamics constraints.

It is difficult to see any way of avoiding the bounds from Galactic dynamics, so we look to loosen the limits derived in Section~\ref{sec:earlyuniverse}. Clearly if the interaction responsible for freezing out the relic density is not the $U(1)_D$ constrained by soft scattering, then $\hat{\alpha} \lesssim 10^{-3}$ is not ruled out. We consider such examples in the next Section. However, we first consider the possibility that our assumptions in deriving the relic density are too conservative.

From Eq.~(\ref{eq:OmegaDM}), if we reduce $\hat{\alpha}$ (and therefore $\sigma_0$) in order to satisfy the scattering bounds, we must either decrease $x_f$ or increase $g_{*\rm S}/\sqrt{g_*}$. In lowering $\hat{\alpha}$ by a minimum of two orders of magnitude, $x_f/(g_{*\rm S}/\sqrt{g_*})$ must likewise increase. As $x_f$ depends only logarithmically on $\hat{\alpha}$ and the number of degrees of freedom, so it is hard to see how it it could be increased sufficiently to counterbalance $\hat{\alpha}$ of order $10^{-3}$ (rather than $\hat{\alpha} \sim 10^{-2}$). We conclude that the number of effective degrees of freedom must be increased. From Eqs.~(\ref{eq:gstarxi}) and (\ref{eq:gstarSxi}), we see that if $\xi=1$, then at freeze-out we must have
\begin{equation}
\frac{g_{*\rm S}}{\sqrt{g_*}} = \sqrt{\sum_{i = {\rm bosons}} g_i + \frac{7}{8} \sum_{i = {\rm fermions}} g_i} \sim 10^2. \label{eq:dofrestriction}
\end{equation}
From Eq.~(\ref{eq:BBNbound2}), these $\sim 10^4$ degrees of freedom must exist in the visible sector at $T_f$, rather than the dark sector.

Alternatively, we could imagine that there are no (or few) new particles beyond the minimum $\chi$ and $\hat{\gamma}$ at freeze-out, but instead $\xi \gg 1$. In this limit
\begin{equation}
\frac{g_{*\rm S}}{\sqrt{g_*}} \approx \xi \sim 10^2. \label{eq:dofxi}
\end{equation}
This limit is more troublesome; from Eq.~(\ref{eq:BBNbound2}) we saw that large values of $\xi$ at the reheating scale (and subsequently $T_f$) very quickly violate the bounds on relativistic degrees of freedom at BBN. Clearly, by increasing the number of degrees of freedom in the visible sector, this bound could be avoided as well. However, we are left with the conclusion that either $\xi(T_f) \sim 10^2$ or there exist $\sim 10^4$ new particles at a few hundred GeV to a TeV. We leave it to the reader to decide how palatable these alternatives are.

A separate, but conceptually similar, bound on scattering can be placed by considering the interaction of galactic dark matter with the hotter DM of the surrounding cluster. Scattering will cause heating in galactic DM, and eventually evaporate the halo. From Ref.~\cite{Gnedin:2000ea} the characteristic time for this evaporation is given by
\begin{equation}
t_{\rm evap.} = 3.5 \times 10^9 ~\mbox{years} \left(\frac{\sigma/m_\chi}{\mbox{cm$^{2}$/g}} \right)^{-1} \left(\frac{v_{\rm cluster}}{10^3 \mbox{km/s}} \right)^{-1}\left(\frac{\rho_{\rm cluster}}{ 1.3 \times 10^{-3} M_\odot \mbox{pc}^{-3}} \right)^{-1}. \label{eq:evap}
\end{equation}
We may estimate the cross-section for soft-scattering by calling the path length $\lambda$ over which a single particle loses of order its initial kinetic energy $(\Delta v^2/v^2)^{-1}R$, where $R$ is the radius of the galaxy, and $\Delta v^2/v^2$ from Eq.~(\ref{eq:totaldeltav2}) is the fractional energy loss as the particle travels once through the halo. This can be expressed as an effective scattering cross section by setting $\lambda = (n \sigma)^{-1}$, where $n = N/R^3$ is the number density of DM in the halo, we find
\begin{equation}
\frac{\sigma}{m_\chi} \approx \frac{8 \hat\alpha^2}{m_\chi^3 v^4} \ln \left( \frac{GN m_\chi^2}{2 \hat\alpha}\right). \label{eq:sigmam}
\end{equation}
Letting the cluster velocity and density take on the canonical values ($v_{\rm cluster}=10^3 \mbox{km/s}$ and $\rho_{\rm cluster} = 1.3 \times 10^{-3} M_\odot \mbox{pc}^{-3}$, where $M_\odot$ is the solar mass), we can place limits on $\hat\alpha$ and $m_\chi$ by requiring that $t_{\rm evap.}$ is longer than the age of the universe. Numerically, we find this bound less stringent than that from soft-scattering of particles within the Galactic halo, Eq.~\ref{eq:tausoft}.

It is interesting to note that, aside from logarithmic enhancements, the bound placed on $\hat\alpha$ vs.~$m_\chi$ parameter space from soft scattering is essentially a line of constant $\sigma/m_\chi$ (that is, they are, up to log corrections, lines of slope $2/3$ on the log-log plot). As mentioned, limiting DM to one hard scattering in the lifetime of the universe is equivalent to bounding $\sigma/m_\chi$ in the Galaxy to be $\lesssim 0.1~\mbox{cm$^2$/g}$. It has been suggested in the literature that values of $\sigma/m_\chi$ in the range $0.01 - 5~\mbox{cm$^2$/g}$ \cite{MiraldaEscude:2000qt,Wandelt:2000ad,Dave:2000ar,Yoshida:2000uw} may provide better agreement between simulation and observation. Therefore, our limit from soft-scattering should be considered as the general region at which interaction effects may become relevant. Additionally, from Eq.~(\ref{eq:sigmam}) as $\sigma/m_\chi \propto v^{-4}$, it should be expected that the soft-scattering bound will vary greatly in DM systems with a range of virial velocities $v$. In particular, we surmise that a bound even stronger than that estimated here can be obtained from the dwarf galaxies that exhibit the highest observed dark-matter phase-space densities \cite{Dalcanton:2000hn}.

\section{Weakly Coupled Models \label{sec:SU2}}

In this Section, we examine an expanded version of our minimal model: one in which the $\chi$ dark matter particles possess $SU(2)_L$ quantum numbers in addition to a $U(1)_D$ charge.  For such $SU(2)_L \times U(1)_D$ particles, the cross section for freeze-out in the early universe is dominated by the weak interaction $\sigma \sim \alpha^2/m_\chi^2$, and the $U(1)_D$ contribution is negligible for the small values of $\hat{\alpha}$ under consideration. At late times the situation is reversed. The weak cross section remains small, as it is the result of a short-range force. However the long range cross section for soft scattering increases as the dark matter cools and slows, as exemplified in Eq.~(\ref{eq:totaldeltav2}). This allows the strength of $\hat{\alpha}$ to be $\sim 10^{-3}$ as required by Galactic dynamics without running afoul of the relic density conditions, which would require $\hat{\alpha} \sim 10^{-2}$ (when $m_\chi \sim 1$~TeV).

We therefore take our Dirac fermion $\chi$ to be a $({\bf 1},{\bf n})_{Y,D}$ multiplet of $SU(3)_C\times SU(2)_L \times U(1)_Y \times U(1)_D$, where we shall take the $U(1)_D$ coupling to be in the region of Fig.~\ref{fig:alphamass} allowed by soft scattering. Thus  $\hat{\alpha} \lesssim 10^{-3}$. The behavior of this model in the early universe is very similar to the `minimal dark model' of Ref.~\cite{Cirelli:2005uq}, from which we take many of our constraints. 

In outlining our original model in Section~\ref{sec:earlyuniverse}, we set the coefficient of the mixing term $F_{\mu\nu}\hat{F}^{\mu\nu}$ to zero at the high scale. Clearly loops involving $\chi$ would generate a non-zero mixing if the $\chi$ field possesses non-zero hypercharge $Y$. In order to avoid this complication, we set $Y=0$. 

Our $\chi$ particle must be neutral under $U(1)_{EM}$. 
With the assumption of $Y=0$, this requires $\chi$ to sit in an $n$-plet of $SU(2)_L$ where $n$ is odd ({\it i.e.}~$n=3,5,\ldots$). In the spirit of simplicity we take $n=3$, so the $\chi$ triplet contains the neutral $\chi^0$ and (electromagnetically) charged $\chi^\pm$, all with $U(1)_D$ charges of $+1$. Due to $SU(2)_L$ loops, the $\chi^\pm$ are $166$~MeV heavier than the $\chi^0$, and decay before BBN. If the dark matter mass is $m_\chi = 2.4$~TeV, the correct dark matter abundance (including production and then decay of $\chi^\pm$) results from thermal freeze out (see Ref.~\cite{Cirelli:2005uq}). We note that our model does have the nice feature of automatically suppressing unwanted decays of $\chi$ into SM particles, as by assumption $\chi$ is the lightest particle charged under $U(1)_D$.

This minimal model is anomaly free. Triangle diagrams with one or three $SU(2)_L$ vertexes vanish by the tracelessness of the $SU(2)_L$ generators. The diagrams consisting of an odd number of $U(1)_D$ vertexes also vanish as the dark sector contains only two Weyl fermions, one with $+1$ under $U(1)_D$, and the other with $-1$. 

This model does not run afoul of BBN (or CMB) bounds. As in the pure $U(1)_D$ theory, the only new relativistic degrees of freedom at BBN are the two from the $\hat{\gamma}$. Due to the interactions between $\chi$ and the weakly charged SM fields, we expect the temperatures $T$ and $\hat{T}$ to track, so $\xi=1$ until the $\chi$ freeze-out. With small values of $\hat{\alpha}$, the dark photons may freeze-out earlier, and would thus be colder. However, if we take the worse-case scenario that the dark photons do not decouple until after the $\chi$ undergo freeze-out we find (from Eq.~(\ref{eq:BBNbound2})) that BBN bounds are satisfied as long as freeze-out occurs when
\begin{equation}
g_{\ast\rm{vis}} \geq 18.8. \label{eq:weakBBN}
\end{equation}
This is easily satisfied for any model that freezes out before the QCD phase transition.

\begin{figure}[htbp]

\centering

\includegraphics[width=0.3\columnwidth]{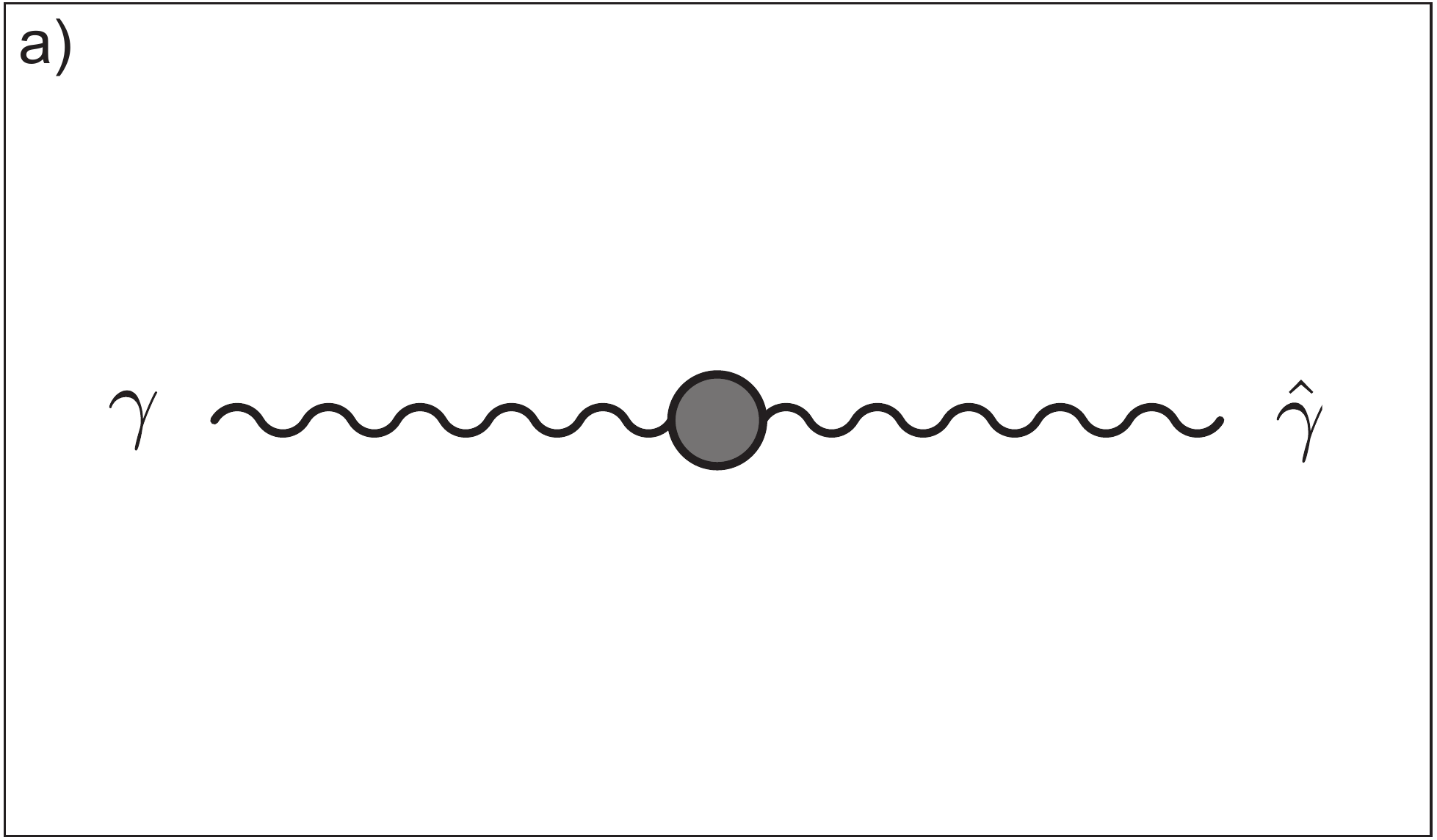}\includegraphics[width=0.3\columnwidth]{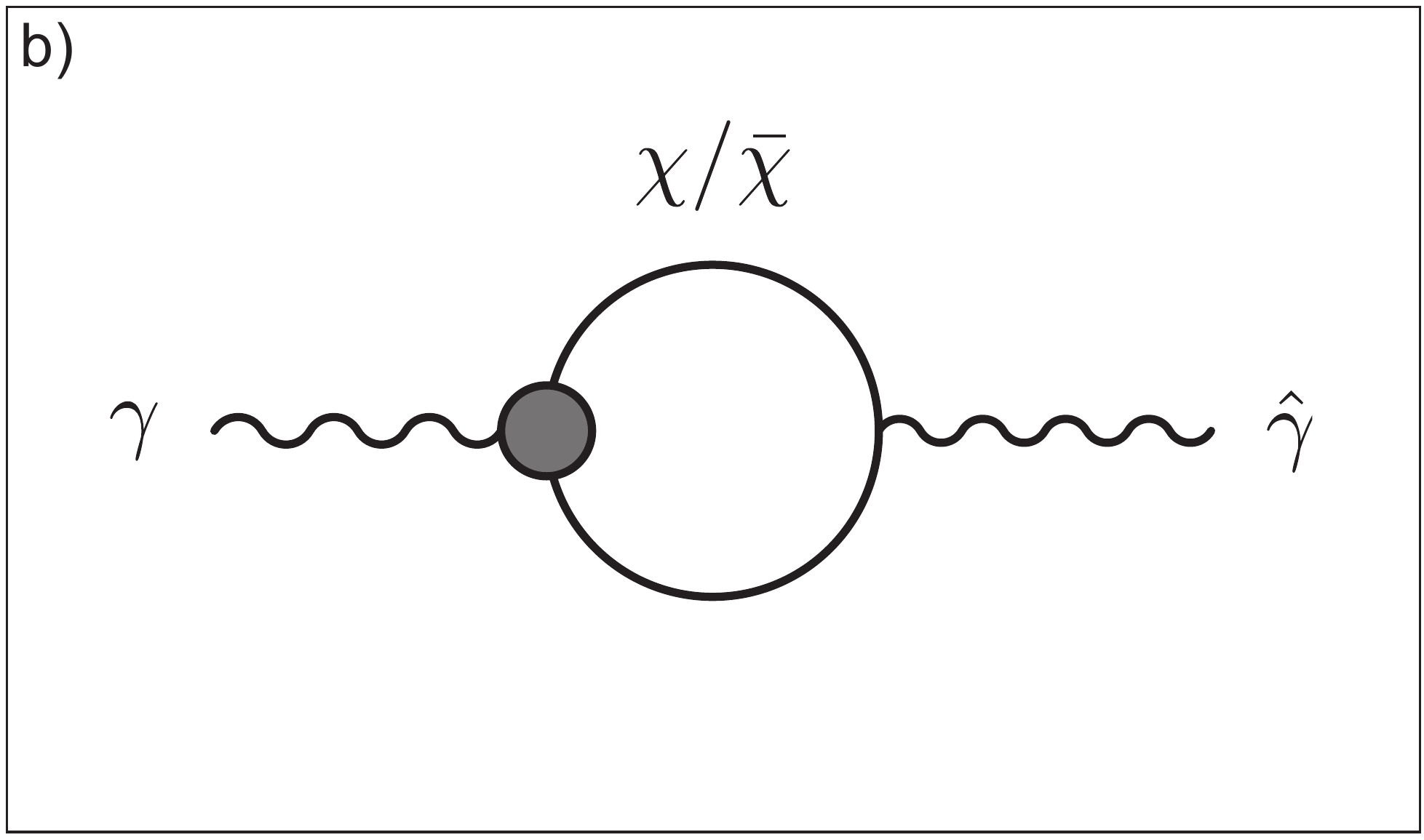}

\caption{Feynman diagrams leading to $\gamma/\hat{\gamma}$ mixing. The vertex in a) can be expanded into that shown in b), as the only particle to which the $\hat{\gamma}$ couples is $\chi/\bar{\chi}$. Since the mass and $SU(2)_L$ charge of these two particles are the same, yet they possess opposite $U(1)_D$ charge, the sum of the $\chi$ and $\bar{\chi}$ diagrams in b) is zero, and the overall mixing vanishes. \label{fig:ghatg}}
\end{figure}

Next we must check that our $\chi$ does not have too large of a coupling to SM particles. We first demonstrate that no mixing occurs between the photon and the dark photon $\hat{\gamma}$. As indicated previously, we assume that there is no $F_{\mu\nu}\hat{F}^{\mu\nu}$ term at high energies. With purely $SU(2)_L\times U(1)_D$ coupling, we find that the diagram Fig.~\ref{fig:ghatg}a vanishes. This is because any such vertex can be rewritten as the $\hat{\gamma}$ coupling to a $\chi$ or $\bar{\chi}$  which then couples to the $\gamma$ through some vertex involving SM fermions and $SU(2)_L$ couplings (Fig.~\ref{fig:ghatg}b). However, since the mass and $SU(2)_L$ couplings of $\chi$ are the same as those of $\bar{\chi}$ yet the $U(1)_D$ charge is opposite, the sum of the two diagrams is zero. 

\begin{figure}[htbp]

\centering

\includegraphics[width=0.3\columnwidth]{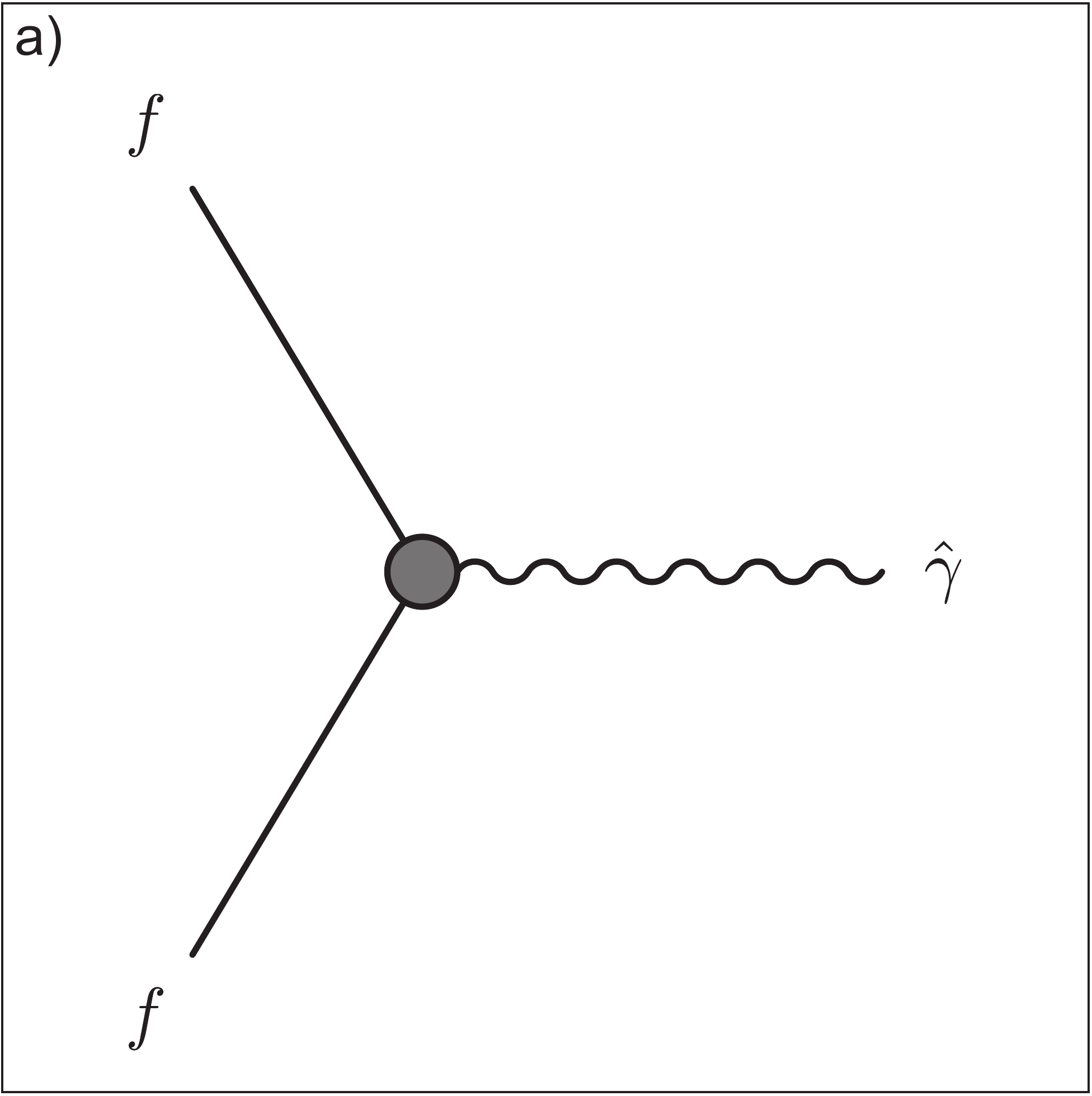}\includegraphics[width=0.3\columnwidth]{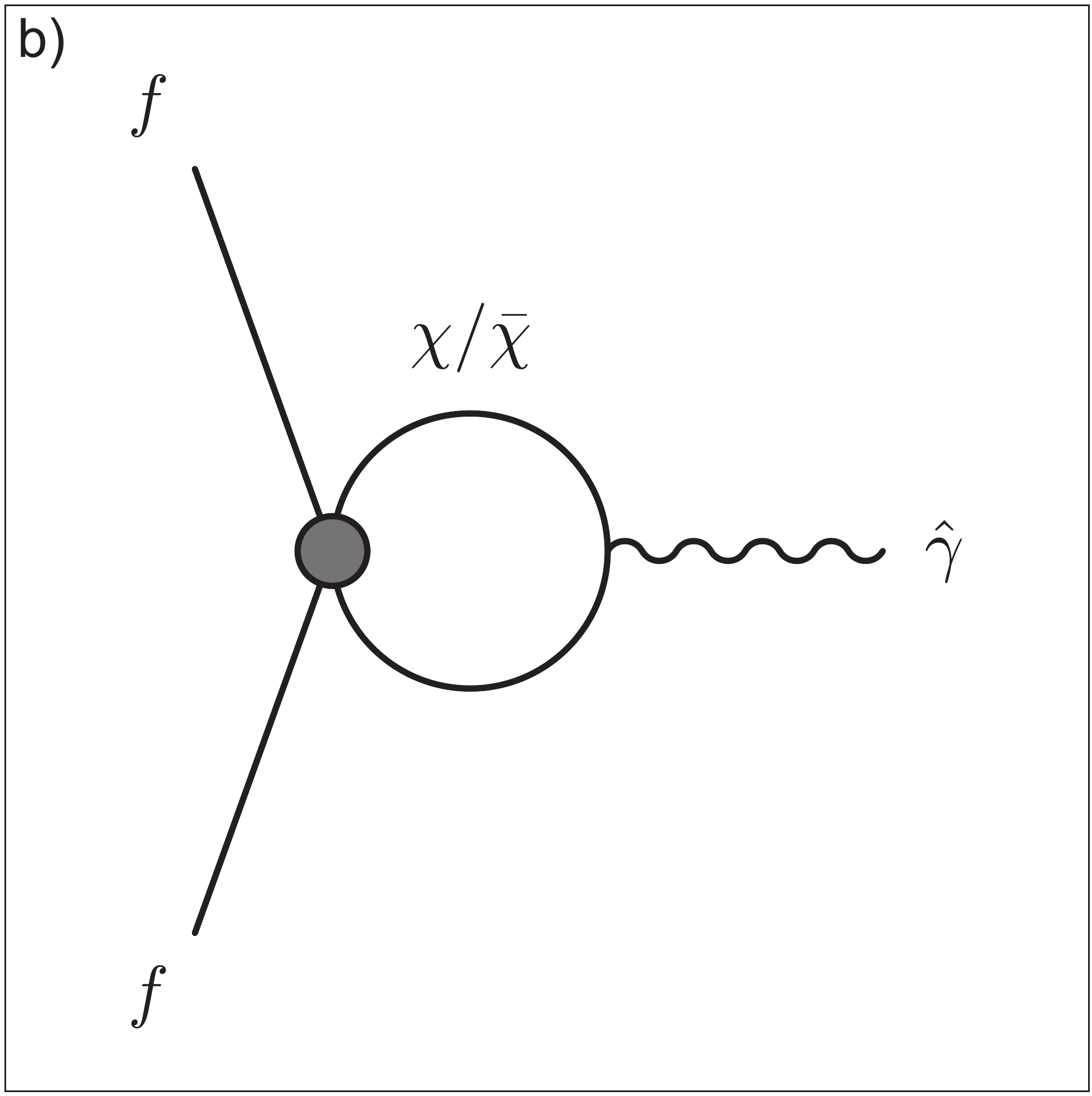}

\caption{Feynman diagram leading to $\hat{\gamma}$ interactions with SM fermions $f$. The vertex in a) can be expanded into that shown in b), as the only particle with an interaction with $\hat{\gamma}$ is the $\chi/\bar{\chi}$. Since the mass and $SU(2)_L$ charge of these two particles are the same, yet the $U(1)_D$ charges are opposite, the sum of the $\chi$ and $\bar{\chi}$ diagrams in b) is zero, and the overall coupling of $f$ to $\hat{\gamma}$ is therefore zero as well. \label{fig:ffhatg}}
\end{figure}

Similarly, the coupling between $\hat{\gamma}$ and a standard model fermion $f$ is also zero. The relevant diagrams are shown in Fig.~\ref{fig:ffhatg}. Again, the vertex between $f$ and $\hat{\gamma}$ (Fig.~\ref{fig:ffhatg}a) can be divided into the $\chi/\bar{\chi}$ vertex connecting with $\hat{\gamma}$ and a vertex between $\chi/\bar{\chi}$ vertex connecting with $f$ (Fig.~\ref{fig:ffhatg}b). As the latter vertex  is identical for $\chi$ and $\bar{\chi}$ but the former has opposite signs, the overall diagram vanishes.

\begin{figure}[htbp]

\centering

\includegraphics[width=0.3\columnwidth]{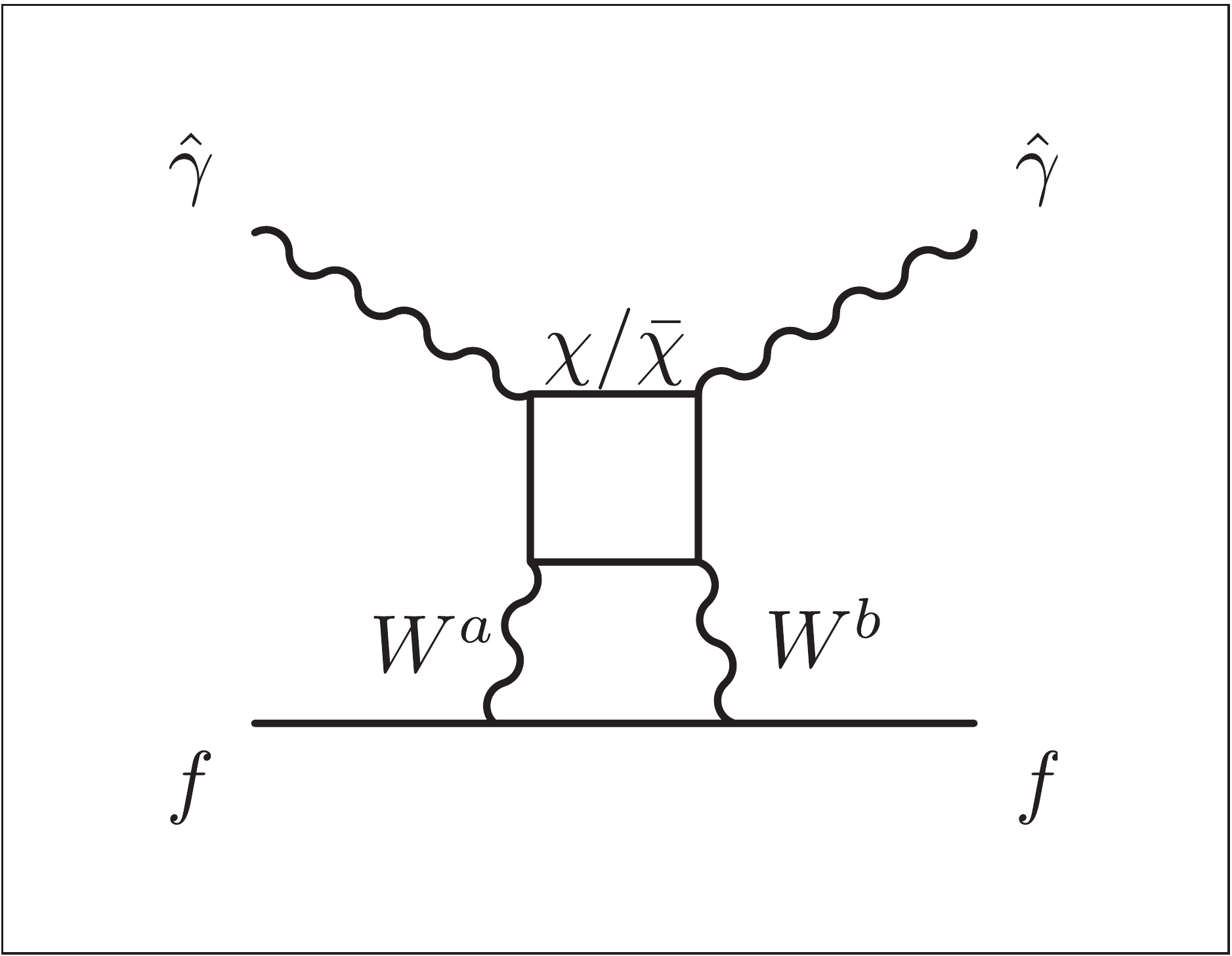}

\caption{The leading order interaction of the dark sector with SM fermions. The dark photons $\hat{\gamma}$ couple to a loop of $\chi$ particles, which couple through two $SU(2)_L$ gauge bosons to SM fermions. Coupling through a single $SU(2)_L$ boson is zero due to the tracelessness of $\tau^a$.
 \label{fig:ffgg}}
\end{figure}

The lowest order coupling of SM fermions to $\hat{\gamma}$ occurs at $\alpha^2\hat{\alpha}$. This is due to a two loop effect, as shown in Fig.~\ref{fig:ffgg}, and unlikely to be accessible in direct detection. We can represent this interaction by an effective Lagrangian whose lowest order term is given by $\frac{\beta}{m_\chi^3}\hat{F}_{\mu \nu}\hat{F}^{\mu \nu} \bar{f}f$ where $\beta=\lambda_{f} \frac{{\alpha}^2 \hat{\alpha}}{4 \pi}$ and $\lambda_{f}$ is the Yukawa coupling of the fermion that is involved.
Let us estimate the order of magnitude of this interaction. To be conservative we use the Yukawa coupling of a $u$ quark and take $\hat{\alpha}=10^{-2}$; which by galactic dynamics is the maximum allowed value for $m_{\chi}\sim 2$~TeV. With these values we find $\beta \sim 10^{-10}$ and $\frac{\beta}{m_{\chi}^3}\sim 10^{-20}\,\rm{GeV}^{-3}$. We estimate that the interaction length for dark photons inside the cores of stars would be on the order of $10^{18}$~km, and thus this interaction would not introduce a potentially dangerous new source of stellar cooling.

Due to the high-order interaction between $\hat{\gamma}$ and SM particles, we cannot expect to directly observe the dark radiation. In addition, while the $\chi$ fields would have a direct detection cross section of $10^{-44}-10^{-45} \mbox{ cm}^2$ \cite{Cirelli:2005uq} and so could be seen in SuperCDMS, any such detection would be indistinguishable from a scenario without the dark photons. Therefore, the presence of a new unbroken $U(1)_D$ in the dark sector could only be probed via its effect on galactic dynamics. Clearly in the limit that $\hat{\alpha}\to 0$, the Galactic structure would remain unchanged. Values of $\hat{\alpha}$ near the maximum allowed from soft-scattering ({\it i.e.}~$\hat{\alpha} \sim 10^{-2}$ for the $SU(2)_L$ triplet candidate with $m_\chi \sim 2$~TeV) should have a measurable effect on the halo structure, as in this regime the dark matter is no longer completely collisionless. A full study of this effect requires simulations beyond the scope of this paper, though some additional considerations are discussed in the following Section.

\section{Other Effects of Dark Photons \label{sec:newlimits}}

The existence of a dark matter `plasma' may have additional effects that could significantly affect structure formation. We mention three possibilities here: bremsstrahlung, early universe structure formation, and the Weibel instability in galactic halos. The first two result in much weaker bounds than those already derived, and are mentioned here only for completeness. The Weibel instability may have significant and visible effects in the halo, but requires simulation beyond the scope of this paper.

\subsection{Bremsstrahlung \label{sec:brem}}
Emission of a soft $\hat{\gamma}$ during a $\chi/\bar{\chi}$ collision could conceivably serve as another energy loss mechanism in the halo on par with soft and hard scattering as outlined in Section~\ref{sec:galacticdynamics}. To derive a bound on $\hat{\alpha}$ as related to $m_\chi$, we make the same assumption as in the case of soft scatter: over the lifetime of the universe, a dark-matter particle cannot lose on order of its initial kinetic energy through bremsstrahlung of dark radiation. By assuming dipole radiation during a soft collision, we find that
\begin{equation}
\frac{3}{64} \frac{G m_\chi^3 R}{ \hat{\alpha}^3} \ln^{-1}\left( \frac{G N m_\chi^2}{2 \hat{\alpha}}\right) \geq 50. \label{eq:brem}
\end{equation}
However this bound is weaker than that from both hard and soft scattering over the parameter space of interest.

\subsection{Structure Formation \label{sec:structure}}

In the early universe, structure cannot grow until after matter/radiation equality. Until the matter (which can clump) decouples from the dark radiation (which cannot), density perturbations remain fixed. We can estimate the scale factor at which this occurs by finding the redshift $z_*$ at which the dissipation time (the time over which the velocity of a dark matter particle is significantly perturbed by the radiation) becomes longer than the Hubble time $H^{-1}$. The argument follows that in Ref.~\cite{Peebles:1994xt} for the decoupling of baryons from the photon bath.

The dissipation time is the logarithmic derivative of the velocity:
\begin{equation}
t^{-1}_{\rm diss} \equiv v^{-1} \frac{dv}{dt} = v^{-1} \frac{F}{m_\chi}\,. \label{eq:tdiss} 
\end{equation}
Here $F$ is the force due to radiation pressure, 
\begin{equation}
F = \frac{4}{3} \hat{\sigma}_T a\hat{T}^4 v\,,  
\end{equation}
where
\begin{equation}
\hat\sigma_T = \frac{8\pi}{3} \frac{\hat\alpha^2}{m_\chi^2}
\end{equation}
is the Thomson cross-section for dark matter interacting with dark photons and (as before) $\hat{T}$ is the temperature of the dark photons. As we shall see, the decoupling occurs when the universe is radiation dominated, so the Hubble time is given by
\begin{equation}
H^2 = \frac{4\pi^3}{45} g_* \frac{T^4}{m_{\rm Pl}^2}. \label{eq:thubble}
\end{equation}
Here $T$ is the photon temperature.

The conservation of entropy relates the photon temperature $T$ at redshift $z_*$ with the photon temperature today, $T_0$,

\begin{equation}
T=\left(\frac{g_{* \,S} (T_0)}{g_{* \,S} (T)}\right)^{1/3} \frac{T_0}{a}. \label{eq:entropy}
\end{equation}

Combining Eqs.~(\ref{eq:tdiss}) and (\ref{eq:thubble}), we find the decoupling redshift $z_*$ to be
\begin{eqnarray}
 1+z_* &=& \frac{3}{16}\sqrt{\frac{\pi}{5}} \xi^{-4}
\frac{m_\chi^3}{\hat\alpha^2 T_0^2 m_{\rm Pl}}
 g_*(T)^{1/2} \left(\frac{g_{*S}(T)}{g_{*S}(T_0)}\right)^{2/3} \nonumber \\
 &=& 2.3 \times 10^{18} \xi^{-4}
\left(\frac{10^{-2}}{\hat\alpha}\right)^2 \left(\frac{m_\chi}{{\rm
TeV}}\right)^3
 g_*(T)^{1/2} \left(\frac{g_{*S}(T)}{g_{*S}(T_0)}\right)^{2/3} \, .
\end{eqnarray}
As before $\xi$ is the ratio of dark photon temperature to photon temperature at redshift $z_*$ (recall that it is difficult to construct models where $\xi$ is much larger than unity). The number of degrees of freedom that contribute to the entropy density today, $g_{* \, S} (T_0)$, is of order unity. The decoupling occurs extremely early, before even dark matter freeze-out.\footnote{This is not a contradiction: freeze-out is the time when the dark particles and antiparticles stop annihilating, while decoupling occurs when the dark photons stop imparting significant velocity to the dark matter.} As a result, it seems that this effect will be cosmologically irrelevant.

\subsection{Plasma Instabilities \label{sec:firehose}}

In Section \ref{sec:galacticdynamics}, we constrained
$\hat\alpha$ by demanding that dark matter be effectively
collisionless in galactic halos, under two-body interactions.
However, there may be collective plasma effects that affect
DM dynamics on timescales much shorter than
those due to two-body interactions.  Unfortunately, it is difficult
to state with confidence what the observational consequences of
those effects will actually be, even if they are relevant.
Given theoretical uncertainties about the nonlinear evolution of such
instabilities, we leave the detailed implications to future work.

As a simple example we consider the Weibel instability \cite{Weibel}, an
exponential magnetic-field amplification that
arises if the plasma particles have an anisotropic velocity distribution.
Such anisotropies could arise, for example, during hierarchical structure
formation as subhalos merge to form more massive halos.  Similar
instabilities in the baryonic gas have been postulated to
account for the magnetic fields in galaxy clusters \cite{Medvedev:2005ep}.
The growth rate $\Gamma$ of the magnetic field is
\begin{equation}
     \Gamma = \omega_p \frac{v}{c} = \sqrt{\frac{(4\pi)^2
     \hat{\alpha} \rho}{m_\chi^2}}
     \frac{v}{c}, \label{eq:firehoseGamma}
\end{equation}
where $\omega_p$ is the plasma frequency, $\rho \approx 0.4~
\mbox{GeV/cm}^3$ is the dark-matter density, and $v$ is the
velocity of the dark matter within the colliding halos. Assuming
$v/c \sim 10^{-3}$, we find
\begin{equation}
     \Gamma \sim 10^{-2} \mbox{s}^{-1} \times \frac{\hat{\alpha}^{1/2}}{(m_\chi
     /\mbox{TeV})}. \label{eq:Gammanumerical}
\end{equation}

To be relevant for galactic-halo formation, the timescale
$\Gamma^{-1}$ for magnetic-field amplification should be shorter
than the dynamical timescale $\tau$ of the merging subhalos.
The instability will be therefore be of interest when
\begin{equation}
     \left(\frac{m_\chi}{\mbox{TeV}}\right) \lesssim 10^{11}
     \hat{\alpha}^{1/2} \left(
     \frac{\tau}{10^6~\mbox{yrs}}\right). \label{eq:firehoselimit}
\end{equation}
This range of $\hat{\alpha}$ and $m_\chi$ encompasses the entire
parameter space of interest for any reasonable value of
$\tau$. Therefore, we suspect that galactic structure will be
affected by plasma effects in the dark matter due to the
$U(1)_D$ even when $\hat{\alpha}$ is not near the boundary of
allowed values from soft scattering.  One possibility is that
nonlinear evolution would result in a strongly magnetized
plasma, and if so, dark matter would be effectively collisional
and thus probably inconsistent with data.  However, theory and
simulations that study the nonlinear evolution of the Weibel
instability for relativistic pair plasmas and nonrelativistic
electron-proton plasmas do not yet agree whether the magnetic
fields survive, and simulations for the equal-mass
nonrelativistic plasma we are considering have not been
performed.  It is therefore premature to conclude that these
instabilities will result in effectively collisional dark
matter; a more detailed study will be required to assess these
effects.

\section{Conclusions \label{sec:conclusion}}

Given how little direct information we have about the nature of dark matter, it is of
crucial importance to explore models in which the DM sector has an interesting
phenomenology of its own.  In many ways, an unbroken $U(1)$ gauge field coupled
to dark matter is a natural way to obtain a long-range interaction between DM particles.
In contrast to the case of hypothetical long-range scalar fields, the
masslessness of the gauge field is protected by a symmetry, and the absence of long-range
violations of the equivalence principle is naturally explained by the overall charge
neutrality of the dark plasma.  New unbroken $U(1)$'s can appear naturally in
unified models.

While a dark $U(1)$ may be realized as a broken symmetry with massive vector bosons, it has been pointed out that there are few constraints on the massless, unbroken case from the early universe. We have verified that the minimal model, with just a single massive Dirac fermion for the dark matter and a massless dark photon, is consistent with limits obtained from the number of relativistic degrees of freedom at BBN, with relatively mild assumptions on the reheating temperature of the dark sector. More complicated models are also allowed, depending on the details of spectrum and reheating.

We found that one cannot build a dark matter model charged under a hidden unbroken $U(1)_D$ in which this new gauge group is responsible for thermal freeze out. As can be seen in Fig.~\ref{fig:alphamass}, the required values of $\hat{\alpha}$ and $m_\chi$ required for the $\chi$ particles to form a thermal relic would violate bounds coming from limits on hard and soft scattering of dark matter in the Galactic halo.  As an important consequence of this argument, models in which 
dark matter couples to an exact copy of ordinary electromagnetism (in particular,
with $\hat\alpha = \alpha$) are ruled out unless $m_\chi >$~a few TeV.
This constrains the parameter space of models with hidden copies of the SM or the MSSM in which the dark matter
is electrically charged, such as the model in Ref.~\cite{Feng:2008mu} where the stau was
suggested as a dark matter candidate.

By adding additional interactions to increase the annihilation cross-section, it is possible to build a scenario with an unbroken dark $U(1)$ and the correct relic abundance. Introducing another short-range force coupling to the $\chi$, for example the familiar $SU(2)_L$, can provide an appropriately large cross section for $\chi/\bar\chi$ annihilation. The new coupling $\hat{\alpha}$ must then be relatively small (compared to the $SU(2)_L$ $\alpha$) in order to evade Galactic dynamics bounds. 

The simplest model which realizes this situation is a Dirac fermion in a triplet of $SU(2)_L$ (in order to avoid $U(1)_Y/U(1)_D$ mixing). Bounds from the early universe then force $m_\chi$ to be on the order of a few TeV, which implies $\hat{\alpha} \lesssim 10^{-2}$. Since all couplings between the dark radiation and the SM enter at two loops (and require two dark photons in the process), it would be very difficult to observe the presence of the new gauge group through direct detection. Instead, the best search strategy would be an indirect one: looking for the effects on Galactic dynamics arising from a soft scattering mediated by a long-range force. Clearly, as $\hat{\alpha}$ goes to zero, the model becomes indistinguishable from minimal weakly coupled dark matter. However, if the coupling is near the limit from soft scattering, one would expect detectable deviations from the assumptions of collisionless dark matter currently used in simulations.

Additionally, since the $U(1)_D$ effectively makes the dark halo a plasma (albeit a 
very cold, tenuous one), there may be other effects on structure formation that constrain this model \cite{Frederiksen:2003qu}. 
We have estimated that the timescale for the Weibel instability in our model is short compared
to relevant timescales for galactic dynamics.  If this instability has a dramatic effect when
subhalos collide during the assembly of a galactic halo, our 
$U(1)_D$ could be excluded for the entire range of interesting parameters. Further work is required to
before we reliably understand the quantitative effects of such instabilities on galactic
dynamics.

This work opens a window to new phenomenological possibilities
within the dark sector.  One avenue for further investigation would
be the possibility of ``dark atoms,'' which would arise if there were
two different stable species with dark charge, each with an asymmetry
in the number density of positive and negative charges (with
one balancing the other to maintain overall charge neutrality).
From there, one is free to contemplate dark chemistry and
beyond.  Dark matter constitutes a large majority of
the matter density of the universe, and there is no reason
to assume {\it a priori} that physics there is any less rich
and interesting than that of ordinary matter.

\begin{acknowledgments}

We are very grateful to Sonny Mantry for collaboration in the early stages of this work,
and to Jonathan Feng, George Field, Josh Frieman, Manoj Kaplinghat, Keith Lee, Arvind Rajaraman, and Mark Wise for helpful comments.  This
work was supported by DoE DE-FG03-92-ER40701 and the Gordon and
Betty Moore Foundation.

\end{acknowledgments}

\end{document}